\documentclass[journal]{vgtc}                   

\graphicspath{{figures/}{pictures/}{images/}{./}} 

\usepackage{times}                     
\usepackage{url}

\usepackage{tabu}                      
\usepackage{booktabs}                  
\usepackage{lipsum}                    
\usepackage{mwe}                       

\usepackage{mathptmx}                  

\onlineid{0}

\vgtccategory{Research}

\newcommand{\eg}{e.g.,}
\newcommand{\ie}{i.e.,}
\newcommand{\etal}{{\em et al.}}

\newcommand{\bpstart}[1]{\vspace{1mm}\noindent{\textbf{#1.}}}

\title{Accessible Data Access and Analysis by \\ People who are Blind or Have Low Vision}

\author{%
  Samuel Reinders,
  Munazza Zaib,
  Matthew Butler, 
  Bongshin Lee,
  Ingrid Zukerman,
  Lizhen Qu, and
  Kim Marriott
}

\authorfooter{
  \item
  	Samuel Reinders, Munazza Zaib, Matthew Butler, Ingrid Zukerman, Lizhen Qu, and Kim Marriott are with Monash University.
  	E-mail: first.last@monash.edu
  \item
  	Bongshin Lee is with Yonsei University,
  	E-mail: b.lee@yonsei.ac.kr
}

\teaser{
  \centering
  \includegraphics[width=\linewidth]{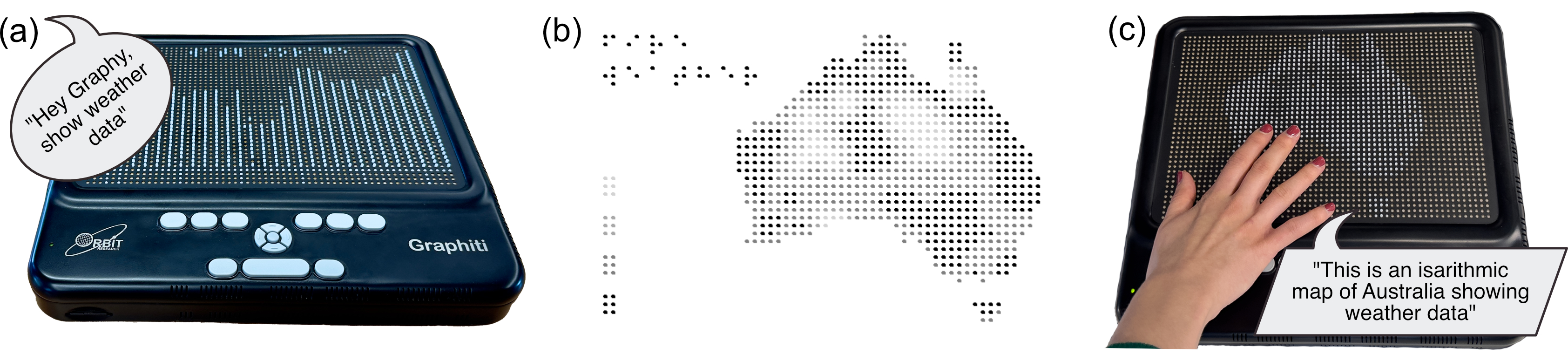}
    \caption{We are exploring how refreshable tactile displays (RTDs) can be combined with conversational agents to assist blind or low-vision (BLV) people in data analysis activities: a)~a bar chart showing water storage over a 27-year period is displayed on the RTD as a user asks the agent to load a new data set; b)~an isarithmic map of Australia based on weather data to be generated on the RTD; c)~as the isarithmic map is loaded onto the RTD, a user begins to touch it, while the agent describes the map.}
    \label{fig:teaser}
}

\abstract{
    Our work aims to develop new assistive technologies that enable blind or low vision (BLV) people to explore and analyze data readily. At present, barriers exist for BLV people to explore and analyze data, restricting access to government, health and personal data, and limiting employment opportunities. This work explores the co-design and development of an innovative system to support data access, with a focus on the use of refreshable tactile displays (RTDs) and conversational agents. The envisaged system will use a combination of tactile graphics and speech to communicate with BLV users, and proactively assist with data analysis tasks. As well as addressing significant equity gaps, our work expects to produce innovations in assistive technology, multimodal interfaces, dialogue systems, and natural language understanding and generation. 
} 


\begin{document}


\maketitle

\section{Introduction}
The last two decades have witnessed a sharp rise in the amount of data available to business, government and science. The proliferation of data and the widespread adoption of data analytics have made data literacy a requisite skill for all professions, not just data analysts or data scientists. However, there is a severe accessibility issue for people who are blind or have low vision (BLV). Textual (or spoken) summaries of data are insufficient, as they do not provide opportunities for independent exploration of the data and validation of findings. Thus, at present, essential data literacy is beyond the reach of most BLV people.

Published guidelines recommend tactile graphics as best practice~\cite{BANA2010guidelines}. Tactile graphics enable BLV people to focus on areas of interest and make their own interpretations, thereby improving levels of agency. Excitingly, \textit{Refreshable Tactile Displays} (\textit{RTDs}\/) have the potential for making tactile graphics delivery easier, faster and more affordable than static tactile graphics. RTDs consist of a grid of pins controlled by actuators to facilitate the presentation of tactile graphics almost instantaneously. Until recently, RTDs large enough for a tactile graphic were prohibitively expensive, and there has been little research into their use. However, the Monarch~\cite{APH}, DotPad~\cite{DotInc} and Graphiti~\cite{Orbit} (Figure~\ref{fig:teaser}) RTDs are coming to market at more affordable price points.

The widespread adoption of conversational agents, particularly among the BLV community, also offers opportunities for supporting data access and analysis. Our work seeks to co-design and implement an intelligent multimodal system that combines conversational AI with tactile presentation and interaction to enable BLV people to explore and analyse data. Such a system could respond to queries and proactively guide users based on its analysis of the data and users' actions. 

\section{Related Work}
\subsection{Accessible Data visualization}
The growing availability of data has allowed decision-making to become increasingly data-driven in virtually all professions. The use of data visualization can, however, be problematic for BLV people due to its graphical nature. Therefore, we are witnessing a surge of research interest in investigating accessible alternatives~\cite{lee2020reaching,kim2021accessible,marriott2021inclusive,dagstuhl2023inclusive}.

One approach is to include a written or spoken description of a graphic. These capabilities have been integrated into systems designed to support data analysis by BLV people~\cite{SharifEtAlCHI2022,alam2023seechart,Kim2023}. Other work has explored the use of sonification (non-speech audio) to convey information, which can be used as an alternative format to data visualizations or alongside data visualizations~\cite{Kramer2010}. Recently, there has been interest in combining speech with sonification~\cite{holloway2022infosonics,thompson2023chart}.

Traditionally speaking, many BLV students at school encounter raised line drawings, known as tactile graphics. Transcription guidelines recommend the use of tactile graphics for accessible provision of maps, diagrams and graphs, rather than textual or verbal descriptions~\cite{BANA2010guidelines}. While most tactile graphics are produced using swell paper or embossing~\cite{rowell2003world}, printed tactile graphics are not well-suited for interactive data exploration, because of the cost and speed of production.

\subsection{Refreshable Tactile Displays}
Refreshable tactile displays (RTDs) constitute a promising alternative to static tactile displays. Most RTDs consist of a grid of pins controlled by electro-mechanical actuators~\cite{Yang2021}. Their great advantage over traditional tactile graphics is that it takes only a few seconds to display a new graphic. In other words, an RTD is the tactile equivalent of a computer monitor, and like a monitor, it supports interactive data exploration.

The DAISY Consortium's Transforming Braille Project has spurred the development of cheaper RTDs. These include the DotPad~\cite{DotInc} and Monarch~\cite{APH}, which provide 2,400 (60$\times$40) and 3,840 (96$\times$40) pins respectively. The Graphiti and Metec~\cite{Metec} offer 2,400 and 6,240 pins, but with several pin heights.

Research into the use of RTDs has focused on the display of static images (\eg\ art and illustrations~\cite{Namdev2015}), dynamic images (\eg\ sports~\cite{Ohshima2021football}) and animations~\cite{Holloway2022}, as well as maps~\cite{brayda2019refreshable}. Despite the huge potential benefits of RTDs, there has been limited research into the use of RTDs for interactive data visualization. Elavsky~\etal\ co-designed tools for the interactive exploration of set diagrams and parallel vectors on an RTD~\cite{elavsky2023data}, and Holloway~\etal\ examined stakeholder perspectives of how RTDs could be used to improve access to data graphics~\cite{holloway2024refreshable}. Systems have also been devised that pair RTDs with audio labels to display web-based maps that support search, zooming, panning and route finding~\cite{Schmitz2012,zeng2014examples,Zeng2015}.

\subsection{Conversational Agents}
Conversational agents, such as Siri and Alexa, are emerging as a promising paradigm to provide natural and engaging interaction experiences both to the general community and to BLV people~\cite{DBLP:conf/chi/BaezCMPPSCP22}. However, at present, they are limited to Internet searches and pre-programmed activities, such as setting timers and alarms, and playing music or reading audiobooks. Work has begun to explore the pairing of conversational agents with visualizations~\cite{DBLP:conf/uist/GaoDALK15,Mitra2022,Narechania2021,DBLP:conf/chi/FastCMBB18}. These however, do not consider BLV users or RTDs, and do not assist in data analysis.

While it is common to augment tactile graphics with predefined audio labels that can be triggered by touch~\cite{butler2021technology}, to the best of our knowledge, the only other research combining a conversational agent with tactile graphics is Jido~\cite{DBLP:conf/assets/QueroBLLLC19}. In Jido, the graphic is a tactile map that supports navigation between points of interest, but the agent has very limited domain knowledge. We are inspired by enthusiastic reactions to our previous research, which added a simple conversational interface to an accessible 3D printed model of the solar system ~\cite{Reinders2020}. We believe that conversational agents would complement RTDs, allowing BLV users to explore visualizations by querying available information using natural language and explore content rendered on an RTD.

\section{Project Aims}
Our work aims to develop new assistive technologies that will enable BLV people to explore and analyse data more readily. The expected outcome of our project is an innovative multimodal system that combines conversational AI and tactile graphics rendered on an RTD to communicate with a BLV user and proactively assists with data analysis tasks. This should provide significant benefits, as it will overcome barriers to data analysis and exploration by BLV people that currently restrict access to government, health and personal data, and limit employment opportunities. Our main aims include:
\begin{itemize}
    \item Identifying how BLV users want to interact with a conversational agent that can help them explore and analyze data.
    
    \item Determining how to generate multimodal presentations of data using a mix of (dynamic) tactile graphics, natural language and non-speech audio.
    
    \item Exploring novel ways in which the agent can support natural multimodal interaction, using a mixture of touch and speech.

    \item Evaluating whether the system is useful, and determining the roles played by the different modalities and their ensuing benefits.   
\end{itemize}

The combination of an RTD and a conversational agent may overcome many of the current barriers BLV people face when analyzing and exploring data. This combination of modalities, which has not been previously considered for this purpose, is the focus of our work.

\section{Summary of Work To Date}
We ran a Wizard-of-Oz (WOz) study to help build an initial understanding of how BLV people would use and react to such a multimodal system~\cite{Reinders2024}. In our study, involving 11 participants, we told the participants that they could use direct touch input and gestures to control the RTD, and engage in speech interactions with a virtual assistant (a text-to-speech interface managed by the wizard). Participants completed several tasks that focused on basic data understanding and analysis, across different scenarios. We presented time series data using line charts and bar charts, and spatial data using isarithmic maps.

The main findings of our WOz study included:
\begin{itemize}
    \item Almost all of our participants felt that using a combination of an RTD and conversational agent to analyse data had benefits over single formats (\ie\ RTD only or agent only) or traditional tactile graphics alone.
    
    \item Depending on the task type, participants gravitated towards different patterns of interactions. For example, touch was crucial during initial exploration, while gestures and speech were more frequently used when identifying values or extrema.
    
    \item Participants' level of tactile experience influenced their interaction patterns, and more experienced tactile graphic users spoke of how the RTD enabled a deeper engagement with the data and independent interpretation.
\end{itemize}

Our findings highlight the desire by BLV people to engage in independent and meaningful data exploration, and indicate that this can be supported by the combination of an RTD and conversational agent. This work has been accepted for publication at IEEE VIS 2024.

Going forward, we are conducting a series of co-design workshops with two BLV co-designers to develop a working prototype system. The focus of these sessions has been on:
\begin{itemize}
    \item \textbf{Understanding the role of the conversational agent.} We are attempting to understand different conversation patterns, the nature of questions asked by users (\eg\ value retrieval, definition questions, Yes/No questions, or follow-up questions), and information needs that can be answered using just speech output or speech along with other modalities.
    
    \item \textbf{Exploring mixed modality interactions.} We are focusing on understanding how multiple modalities, \eg\ touch and voice, can be used together to allow more natural inputs, and how they can be coalesced to deliver more meaningful outputs and responses.

    \item \textbf{Different visualization types and tactile representations.} We are eliciting display options of interest to our co-designers, and collaborating with them to determine whether these charts can be effectively rendered for touch within the constraints of RTDs.
\end{itemize}

\begin{figure}[h]
   \centering
 \includegraphics[width=.85\columnwidth]{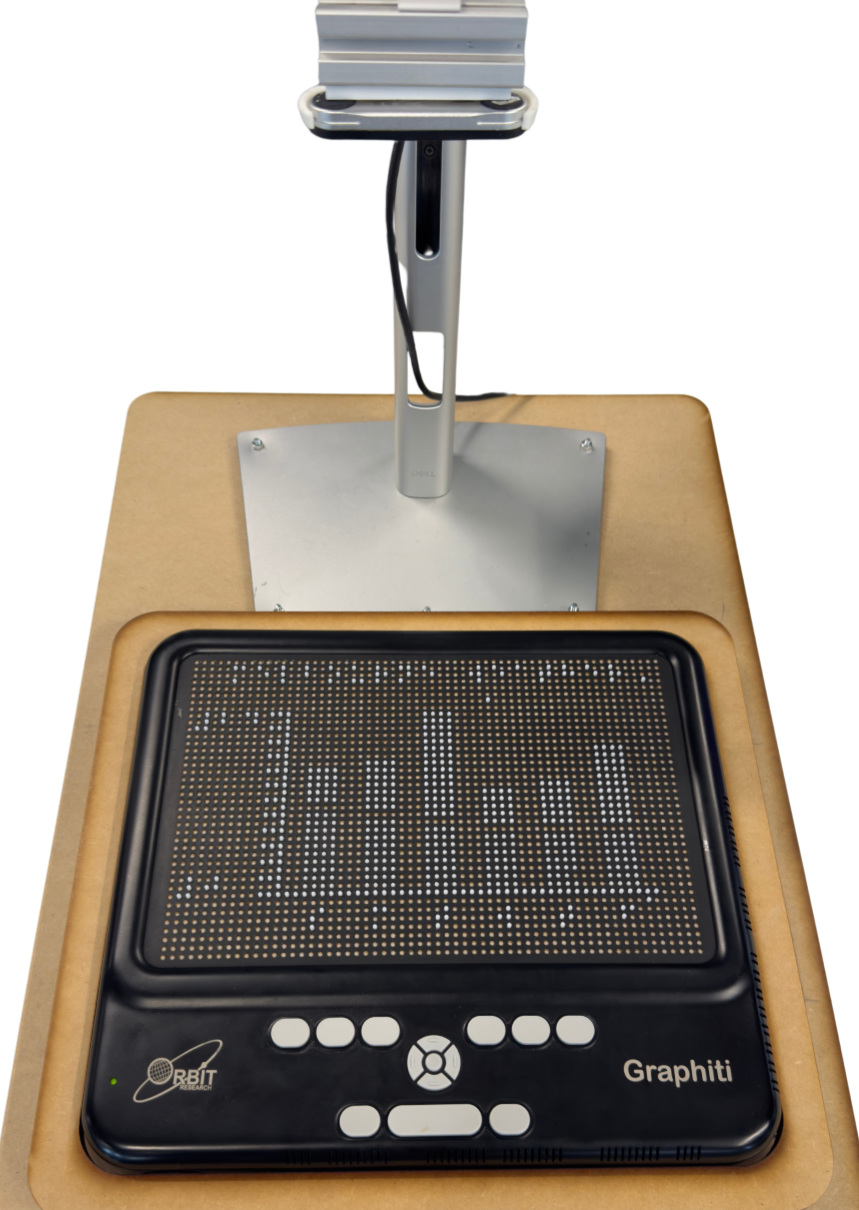}
 \caption{The Graphiti RTD is placed in a stand with a monitor arm that holds the Ultraleap LMC.}\label{fig:graphiti}
\end{figure}

\begin{figure}[h]
   \centering
 \includegraphics[width=0.75\columnwidth]{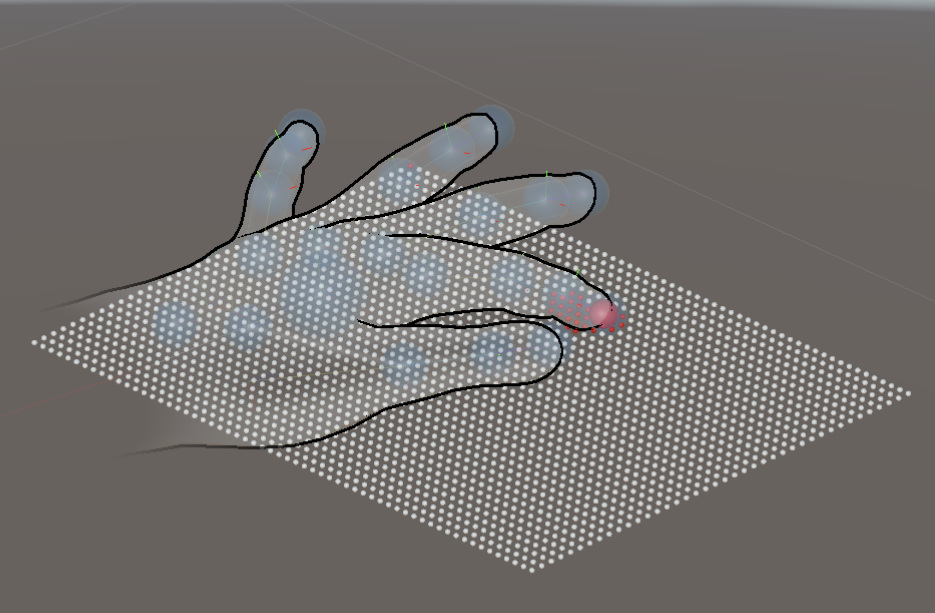}
 \caption{The Ultraleap LMC tracks touch input on the RTDs surface.}\label{fig:touch}
\end{figure}

Based on our WOz study and our initial co-design sessions, we are implementing a multimodal system for data access. Our current prototype (Figure~\ref{fig:graphiti}) utilises the Graphiti RTD and the RASA conversational AI platform (\url{rasa.com}). The Ultraleap Leap Motion Controller (\url{ultraleap.com}) is being used to provide single point touch input (Figure~\ref{fig:touch}), and in the future, we intend to train a gesture model to recognize dynamic touch gestures like pinching and swiping. We plan to conduct further co-design sessions to refine the system design before conducting user testing to evaluate its use and effectiveness in supporting BLV users in data analysis and exploration. We are also exploring how our work can be deployed across other RTDs, including the DotPad and Monarch.

\section{Key Research Challenges}
Our initial findings highlight a desire by BLV people to engage in meaningful exploration of data, and indicate that combining an RTD with a conversational agent is a promising approach. However, there is much to investigate, with key challenges including:

\subsection{Refreshable Tactile Displays}

\vspace{-1mm}\bpstart{Effective presentation of visualizations on RTDs} It is not known how useful current design guidelines regarding tactile graphics are when it comes to rendering graphics on RTDs. Due to the limited display resolution of RTDs, design choices that distil a visualization down to its core components must be made. This may involve a compromise between chart readability, simplification and accuracy. Trials must be undertaken to identify visualization types that are suited to RTDs and to determine how traditional visualization design principles can be adapted accordingly.

\bpstart{The technical limitations of existing RTDs} Current RTDs offer resolutions between 2,400 (60*40) and 3,840 (96*40) pins. In addition, the refresh rates of RTDs are slow, often taking up to 5 seconds for an entire display. As an RTD is effectively a display that is designed to be touched, rather than looked at, these limitations place constraints on the complexity of the graphics that can be rendered on them, and on the naturalness of the interactions with them. Furthermore, our initial work has identified that BLV users would like to interact and manipulate data using direct touch input, \eg\ touch gestures like pinching and swiping, in order to zoom or pan. However, many RTDs either support no direct touch input (\eg\ DotPad) or operate using single-point touch input (\eg\ Monarch, Graphiti). 
 
\bpstart{Supporting interactions like zooming, panning, filtering, search and brushing} Because RTDs allow the display of dynamic content that users can interact with, we must ensure that operations are implemented in ways that support the BLV users' understanding and control of the RTD. This is especially true for operations that require large portions of the RTD to be refreshed (\eg\ zooming and panning), which can cause context loss and user confusion~\cite{zeng2014examples,Zeng2015}. In such cases, additional system outputs may be required to spatially situate a user and ensure that positional context is maintained, \eg\ scroll bars, mini-maps, actuating pins or auditory output.

\subsection{Conversational Agents}

\vspace{-1mm}\bpstart{Handling ambiguities and co-references in conversational queries} Conversational agents may come across ambiguous or inaccurate requests --- an ambiguous request has several interpretations, while an inaccurate request matches several interpretations partially. In addition, ambiguous or inaccurate requests may result from different parse trees or from misheard utterances in spoken interactions. Conversational agents must be able to cope with such issues, \eg\ by reducing the number of options by taking advantage of contextual information~\cite{DBLP:conf/sigir/ChristmannRW22}, or asking clarification questions~\cite{DBLP:conf/emnlp/LeeK0LPLJ23}.

\bpstart{The role of Large Language Models (LLMs)} The increasing pervasiveness of LLMs raises issues of trust and reliability. While LLMs are well suited to handling open-domain conversations, and capable of generating rich responses, incorrect or misleading information resulting from hallucinations can be problematic. This may be especially true for BLV users, who may have a greater reliance on these models than sighted users, because of the difficulty of verifying the provided information. Rather than employing LLMs to deliver accurate information, we envisage using them for language-related tasks, such as paraphrasing users' input and expanding our system's vocabulary.

\subsection{Multimodal Interactions}

\vspace{-1mm}\bpstart{How to best answer questions with a mixed modality system} Combinations of tactile and auditory outputs may assist BLV users in overcoming the loss of bandwidth caused by vision loss. Further investigation is required to identify when and how to effectively combine these modalities, \eg\ search could be supported using pin actuation, where data points referenced in an agent's response could be actuated in real-time on the RTD. We must also identify situations and question types where outputs may be better delivered in a unimodal manner, \eg\ a mean value may be effectively communicated through speech, or a trend line rendered on the RTD for touch reading.

\bpstart{How to handle seamless and natural multimodal interactions} There are a few problems in multimodal interactions that are exacerbated for BLV users. For instance, systems should be highly tolerant of errors, and provide an easy way to correct mistakes or repeat previous steps without having to start over; and they must be contextually aware of a user's input, both touch and speech, and their combination (\eg\ using demonstrative pronouns, such as `this' and `those'). To do so, we must deal with uncertainty in the location of a display touch and with multiple fingers touching the display.

\subsection{System Usage}

\vspace{-1mm}\bpstart{Dealing with individual needs, preferences, and desired independence} Each user brings their own unique set of experiences and preferences that influence how they might choose to interact with the envisaged system to achieve their informational needs~\cite{Reinders2023,Choi2020,Abdolrahmani2018}. Certain users may have more tactile reading experience, and thus desire independent touch-based exploration, whereas other users may rely more heavily on natural language. Data literacy plays a role too, as different users may have different levels of experience with different types of visualizations, and require support from the conversational agent accordingly. Levels of support should be configurable by users, allowing them to set boundaries lie that determine independence of interpretation.
  
 \bpstart{Application areas of such a data access device} We have identified several key areas of everyday life where BLV users desired more equitable access to data. This includes education (\eg\ statistics, mathematics, computing algorithms), workplace (\eg\ stock analysis, maps), and personal data (\eg\ personal finance, health, sports and weather). Further investigation of these use cases is required, identifying how such a device can be tailored to better support informational needs specific to each of these application areas, all the while respecting the needs and preferences of the user.

\section{Conclusion}
We assert that the combination of RTDs and conversational agents offers compelling benefits for supporting the data access needs of BLV users, and that it presents an exciting and important avenue of research. We invite the data visualization community to share their expertise and to join us on this journey.


\bibliographystyle{abbrv-doi-hyperref}

\bibliography{references}

\begin{thebibliography}{10}

\bibitem{Abdolrahmani2018}
A.~Abdolrahmani, R.~Kuber, and S.~M. Branham.
\newblock "siri talks at you": An empirical investigation of voice-activated personal assistant (vapa) usage by individuals who are blind.
\newblock In {\em Proc. ACM SIGACCESS Conference on Computers \& Accessibility}, ASSETS '18,  10 pages, pp. 249--258. ACM, New York, 2018. \href{https://doi.org/10.1145/3234695.3236344}
{doi: {{%
10\hspace{.1pt}\discretionary{.}{%
}{.}\hspace{.4pt}1145\discretionary{/}{%
}{/}3234695\hspace{.1pt}\discretionary{.}{%
}{.}\hspace{.4pt}3236344}}}


\bibitem{alam2023seechart}
M.~Z.~I. Alam, S.~Islam, and E.~Hoque.
\newblock Seechart: Enabling accessible visualizations through interactive natural language interface for people with visual impairments.
\newblock In {\em Proc. International Conference on Intelligent User Interfaces}, IUI '23,  19 pages, p. 46–64. ACM, New York, 2023. \href{https://doi.org/10.1145/3581641.3584099}
{doi: {{%
10\hspace{.1pt}\discretionary{.}{%
}{.}\hspace{.4pt}1145\discretionary{/}{%
}{/}3581641\hspace{.1pt}\discretionary{.}{%
}{.}\hspace{.4pt}3584099}}}


\bibitem{APH}
American Printing House for the Blind.
\newblock {\em Meet Monarch}, 2023.

\bibitem{DBLP:conf/chi/BaezCMPPSCP22}
M.~B{\'{a}}ez, C.~M. Cutrupi, M.~Matera, I.~Possaghi, E.~Pucci, G.~Spadone, C.~Cappiello, and A.~Pasquale.
\newblock Exploring challenges for conversational web browsing with blind and visually impaired users.
\newblock In {\em Proc. CHI Conference on Human Factors in Computing Systems}, pp. 234:1--234:7. {ACM}, 2022. \href{https://doi.org/10.1145/3491101.3519832}
{doi: {{%
10\hspace{.1pt}\discretionary{.}{%
}{.}\hspace{.4pt}1145\discretionary{/}{%
}{/}3491101\hspace{.1pt}\discretionary{.}{%
}{.}\hspace{.4pt}3519832}}}


\bibitem{brayda2019refreshable}
L.~Brayda, F.~Leo, C.~Baccelliere, C.~Vigini, and E.~Cocchi.
\newblock A refreshable tactile display effectively supports cognitive mapping followed by orientation and mobility tasks: A comparative multi-modal study involving blind and low-vision participants.
\newblock In {\em Proc. 2nd Workshop on Multimedia for Accessible Human Computer Interfaces}, MAHCI '19,  7 pages, p. 9–15. ACM, New York, 2019. \href{https://doi.org/10.1145/3347319.3356840}
{doi: {{%
10\hspace{.1pt}\discretionary{.}{%
}{.}\hspace{.4pt}1145\discretionary{/}{%
}{/}3347319\hspace{.1pt}\discretionary{.}{%
}{.}\hspace{.4pt}3356840}}}


\bibitem{butler2021technology}
M.~Butler, L.~M. Holloway, S.~Reinders, C.~Goncu, and K.~Marriott.
\newblock Technology developments in touch-based accessible graphics: A systematic review of research 2010-2020.
\newblock In {\em Proc. CHI Conference on Human Factors in Computing Systems}, CHI '21,  article no. 278,  15 pages. ACM, New York, 2021. \href{https://doi.org/10.1145/3411764.3445207}
{doi: {{%
10\hspace{.1pt}\discretionary{.}{%
}{.}\hspace{.4pt}1145\discretionary{/}{%
}{/}3411764\hspace{.1pt}\discretionary{.}{%
}{.}\hspace{.4pt}3445207}}}


\bibitem{Choi2020}
D.~Choi, D.~Kwak, M.~Cho, and S.~Lee.
\newblock "nobody speaks that fast!" an empirical study of speech rate in conversational agents for people with vision impairments.
\newblock In {\em Proc. CHI Conference on Human Factors in Computing Systems}, CHI '20,  13 pages, p. 1–13. ACM, New York, 2020. \href{https://doi.org/10.1145/3313831.3376569}
{doi: {{%
10\hspace{.1pt}\discretionary{.}{%
}{.}\hspace{.4pt}1145\discretionary{/}{%
}{/}3313831\hspace{.1pt}\discretionary{.}{%
}{.}\hspace{.4pt}3376569}}}


\bibitem{DBLP:conf/sigir/ChristmannRW22}
P.~Christmann, R.~S. Roy, and G.~Weikum.
\newblock Conversational question answering on heterogeneous sources.
\newblock In E.~Amig{\'{o}}, P.~Castells, J.~Gonzalo, B.~Carterette, J.~S. Culpepper, and G.~Kazai, eds., {\em Proc. ACM SIGIR Conference on Research and Development in Information Retrieval}, pp. 144--154. {ACM}, 2022. \href{https://doi.org/10.1145/3477495.3531815}
{doi: {{%
10\hspace{.1pt}\discretionary{.}{%
}{.}\hspace{.4pt}1145\discretionary{/}{%
}{/}3477495\hspace{.1pt}\discretionary{.}{%
}{.}\hspace{.4pt}3531815}}}


\bibitem{DotInc}
Dot Inc.
\newblock {\em Dot Pad}, 2022.

\bibitem{elavsky2023data}
F.~Elavsky, L.~Nadolskis, and D.~Moritz.
\newblock Data navigator: An accessibility-centered data navigation toolkit.
\newblock {\em IEEE Transactions on Visualization \& Computer Graphics}, 30(01):803--813, jan 2024. \href{https://doi.org/10.1109/TVCG.2023.3327393}
{doi: {{%
10\hspace{.1pt}\discretionary{.}{%
}{.}\hspace{.4pt}1109\discretionary{/}{%
}{/}TVCG\hspace{.1pt}\discretionary{.}{%
}{.}\hspace{.4pt}2023\hspace{.1pt}\discretionary{.}{%
}{.}\hspace{.4pt}3327393}}}


\bibitem{DBLP:conf/chi/FastCMBB18}
E.~Fast, B.~Chen, J.~Mendelsohn, J.~Bassen, and M.~S. Bernstein.
\newblock Iris: {A} conversational agent for complex tasks.
\newblock In {\em Proc. CHI Conference on Human Factors in Computing Systems}, p. 473. {ACM}, 2018. \href{https://doi.org/10.1145/3173574.3174047}
{doi: {{%
10\hspace{.1pt}\discretionary{.}{%
}{.}\hspace{.4pt}1145\discretionary{/}{%
}{/}3173574\hspace{.1pt}\discretionary{.}{%
}{.}\hspace{.4pt}3174047}}}


\bibitem{DBLP:conf/uist/GaoDALK15}
T.~Gao, M.~Dontcheva, E.~Adar, Z.~Liu, and K.~G. Karahalios.
\newblock Datatone: Managing ambiguity in natural language interfaces for data visualization.
\newblock In C.~Latulipe, B.~Hartmann, and T.~Grossman, eds., {\em Proc. ACM Symposium on User Interface Software \& Technology}, pp. 489--500. {ACM}, 2015. \href{https://doi.org/10.1145/2807442.2807478}
{doi: {{%
10\hspace{.1pt}\discretionary{.}{%
}{.}\hspace{.4pt}1145\discretionary{/}{%
}{/}2807442\hspace{.1pt}\discretionary{.}{%
}{.}\hspace{.4pt}2807478}}}


\bibitem{Holloway2022}
L.~Holloway, S.~Ananthanarayan, M.~Butler, M.~T. De~Silva, K.~Ellis, C.~Goncu, K.~Stephens, and K.~Marriott.
\newblock Animations at your fingertips: Using a refreshable tactile display to convey motion graphics for people who are blind or have low vision.
\newblock In {\em Proc. ACM SIGACCESS Conference on Computers \& Accessibility}, ASSETS '22,  article no. 32,  16 pages. ACM, New York, 2022. \href{https://doi.org/10.1145/3517428.3544797}
{doi: {{%
10\hspace{.1pt}\discretionary{.}{%
}{.}\hspace{.4pt}1145\discretionary{/}{%
}{/}3517428\hspace{.1pt}\discretionary{.}{%
}{.}\hspace{.4pt}3544797}}}


\bibitem{holloway2024refreshable}
L.~Holloway, P.~Cracknell, K.~Stephens, M.~Fanshawe, S.~Reinders, K.~Marriott, and M.~Butler.
\newblock Refreshable tactile displays for accessible data visualisation.
\newblock {\em IEEE Visualization (VIS)}, 2023. \href{https://doi.org/10.48550/arXiv.2401.15836}
{doi: {{%
10\hspace{.1pt}\discretionary{.}{%
}{.}\hspace{.4pt}48550\discretionary{/}{%
}{/}arXiv\hspace{.1pt}\discretionary{.}{%
}{.}\hspace{.4pt}2401\hspace{.1pt}\discretionary{.}{%
}{.}\hspace{.4pt}15836}}}


\bibitem{holloway2022infosonics}
L.~M. Holloway, C.~Goncu, A.~Ilsar, M.~Butler, and K.~Marriott.
\newblock Infosonics: Accessible infographics for people who are blind using sonification and voice.
\newblock In {\em Proc. CHI Conference on Human Factors in Computing Systems}, CHI '22,  article no. 480,  13 pages. ACM, New York, 2022. \href{https://doi.org/10.1145/3491102.3517465}
{doi: {{%
10\hspace{.1pt}\discretionary{.}{%
}{.}\hspace{.4pt}1145\discretionary{/}{%
}{/}3491102\hspace{.1pt}\discretionary{.}{%
}{.}\hspace{.4pt}3517465}}}


\bibitem{Kim2023}
J.~Kim, A.~Srinivasan, N.~W. Kim, and Y.-S. Kim.
\newblock Exploring chart question answering for blind and low vision users.
\newblock In {\em Proc. CHI Conference on Human Factors in Computing Systems}, CHI '23,  article no. 828,  15 pages. ACM, New York, 2023. \href{https://doi.org/10.1145/3544548.3581532}
{doi: {{%
10\hspace{.1pt}\discretionary{.}{%
}{.}\hspace{.4pt}1145\discretionary{/}{%
}{/}3544548\hspace{.1pt}\discretionary{.}{%
}{.}\hspace{.4pt}3581532}}}


\bibitem{kim2021accessible}
N.~W. Kim, S.~C. Joyner, A.~Riegelhuth, and Y.~Kim.
\newblock Accessible visualization: Design space, opportunities, and challenges.
\newblock {\em Computer Graphics Forum}, 40(3):173--188, 2021. \href{https://doi.org/10.1111/cgf.14298}
{doi: {{%
10\hspace{.1pt}\discretionary{.}{%
}{.}\hspace{.4pt}1111\discretionary{/}{%
}{/}cgf\hspace{.1pt}\discretionary{.}{%
}{.}\hspace{.4pt}14298}}}


\bibitem{Kramer2010}
G.~Kramer, B.~Walker, T.~Bonebright, P.~Cook, J.~H. Flowers, N.~Miner, and J.~Neuhoff.
\newblock Sonification report: Status of the field and research agenda.
\newblock Technical report, Department of Psychology, University of Nebraska, Lincoln, Nebraska, 2010.

\bibitem{lee2020reaching}
B.~Lee, E.~K. Choe, P.~Isenberg, K.~Marriott, and J.~Stasko.
\newblock Reaching broader audiences with data visualization.
\newblock {\em IEEE computer graphics and applications}, 40(2):82--90, 2020.

\bibitem{dagstuhl2023inclusive}
B.~Lee, K.~Marriott, D.~Szafir, and G.~Weber.
\newblock {Inclusive Data Visualization (Dagstuhl Seminar 23252)}, 2024. \href{https://doi.org/10.4230/DagRep.13.6.81}
{doi: {{%
10\hspace{.1pt}\discretionary{.}{%
}{.}\hspace{.4pt}4230\discretionary{/}{%
}{/}DagRep\hspace{.1pt}\discretionary{.}{%
}{.}\hspace{.4pt}13\hspace{.1pt}\discretionary{.}{%
}{.}\hspace{.4pt}6\hspace{.1pt}\discretionary{.}{%
}{.}\hspace{.4pt}81}}}


\bibitem{DBLP:conf/emnlp/LeeK0LPLJ23}
D.~Lee, S.~Kim, M.~Lee, H.~Lee, J.~Park, S.~Lee, and K.~Jung.
\newblock Asking clarification questions to handle ambiguity in open-domain {QA}.
\newblock In H.~Bouamor, J.~Pino, and K.~Bali, eds., {\em Findings of the Association for Computational Linguistics: {EMNLP} 2023, Singapore, December 6-10, 2023}, pp. 11526--11544. Association for Computational Linguistics, 2023. \href{https://doi.org/10.18653/V1/2023.FINDINGS-EMNLP.772}
{doi: {{%
10\hspace{.1pt}\discretionary{.}{%
}{.}\hspace{.4pt}18653\discretionary{/}{%
}{/}V1\discretionary{/}{%
}{/}2023\hspace{.1pt}\discretionary{.}{%
}{.}\hspace{.4pt}FINDINGS\discretionary{%
}{-}{-}EMNLP\hspace{.1pt}\discretionary{.}{%
}{.}\hspace{.4pt}772}}}


\bibitem{marriott2021inclusive}
K.~Marriott, B.~Lee, M.~Butler, E.~Cutrell, K.~Ellis, C.~Goncu, M.~Hearst, K.~McCoy, and D.~A. Szafir.
\newblock Inclusive data visualization for people with disabilities: a call to action.
\newblock {\em Interactions}, 28(3):47–51,  5 pages, apr 2021. \href{https://doi.org/10.1145/3457875}
{doi: {{%
10\hspace{.1pt}\discretionary{.}{%
}{.}\hspace{.4pt}1145\discretionary{/}{%
}{/}3457875}}}


\bibitem{Metec}
Metec.
\newblock {\em HyperBraille}, 2012.

\bibitem{Mitra2022}
R.~Mitra, A.~Narechania, A.~Endert, and J.~Stasko.
\newblock Facilitating conversational interaction in natural language interfaces for visualization.
\newblock In {\em 2022 IEEE Visualization and Visual Analytics (VIS)}, pp. 6--10, 2022. \href{https://doi.org/10.1109/VIS54862.2022.00010}
{doi: {{%
10\hspace{.1pt}\discretionary{.}{%
}{.}\hspace{.4pt}1109\discretionary{/}{%
}{/}VIS54862\hspace{.1pt}\discretionary{.}{%
}{.}\hspace{.4pt}2022\hspace{.1pt}\discretionary{.}{%
}{.}\hspace{.4pt}00010}}}


\bibitem{Namdev2015}
R.~K. Namdev and P.~Maes.
\newblock An interactive and intuitive stem accessibility system for the blind and visually impaired.
\newblock In {\em PETRA: International Conference on PErvasive Technologies Related to Assistive Environments}, pp. 1--7. ACM, 2015. \href{https://doi.org/10.1145/2769493.2769502}
{doi: {{%
10\hspace{.1pt}\discretionary{.}{%
}{.}\hspace{.4pt}1145\discretionary{/}{%
}{/}2769493\hspace{.1pt}\discretionary{.}{%
}{.}\hspace{.4pt}2769502}}}


\bibitem{Narechania2021}
A.~Narechania, A.~Srinivasan, and J.~Stasko.
\newblock Nl4dv: A toolkit for generating analytic specifications for data visualization from natural language queries.
\newblock {\em IEEE Transactions on Visualization and Computer Graphics}, 27(2):369--379, 2021. \href{https://doi.org/10.1109/TVCG.2020.3030378}
{doi: {{%
10\hspace{.1pt}\discretionary{.}{%
}{.}\hspace{.4pt}1109\discretionary{/}{%
}{/}TVCG\hspace{.1pt}\discretionary{.}{%
}{.}\hspace{.4pt}2020\hspace{.1pt}\discretionary{.}{%
}{.}\hspace{.4pt}3030378}}}


\bibitem{BANA2010guidelines}
B.~A. of~North~America.
\newblock {\em Guidelines and Standards for Tactile Graphics}.
\newblock Braille Authority of North America, 2010.

\bibitem{Ohshima2021football}
H.~Ohshima, M.~Kobayashi, and S.~Shimada.
\newblock Development of blind football play-by-play system for visually impaired spectators: Tangible sports.
\newblock In {\em Proc. CHI Conference on Human Factors in Computing Systems}, pp. 1--6. ACM, 2021. \href{https://doi.org/10.1145/3411763.3451737}
{doi: {{%
10\hspace{.1pt}\discretionary{.}{%
}{.}\hspace{.4pt}1145\discretionary{/}{%
}{/}3411763\hspace{.1pt}\discretionary{.}{%
}{.}\hspace{.4pt}3451737}}}


\bibitem{Orbit}
Orbit Research.
\newblock {\em Graphiti}, 2016.

\bibitem{DBLP:conf/assets/QueroBLLLC19}
L.~C. Quero, J.~D.~I. Bartolom{\'{e}}, D.~Lee, Y.~Lee, S.~Lee, and J.~Cho.
\newblock Jido: {A} conversational tactile map for blind people.
\newblock In J.~P. Bigham, S.~Azenkot, and S.~K. Kane, eds., {\em Proc. ACM SIGACCESS Conference on Computers \& Accessibility}, pp. 682--684. {ACM}, 2019. \href{https://doi.org/10.1145/3308561.3354600}
{doi: {{%
10\hspace{.1pt}\discretionary{.}{%
}{.}\hspace{.4pt}1145\discretionary{/}{%
}{/}3308561\hspace{.1pt}\discretionary{.}{%
}{.}\hspace{.4pt}3354600}}}


\bibitem{Reinders2023}
S.~Reinders, S.~Ananthanarayan, M.~Butler, and K.~Marriott.
\newblock Designing conversational multimodal 3d printed models with people who are blind.
\newblock In {\em Proc. 2023 ACM Designing Interactive Systems Conference}, DIS '23,  17 pages, p. 2172–2188. ACM, New York, 2023. \href{https://doi.org/10.1145/3563657.3595989}
{doi: {{%
10\hspace{.1pt}\discretionary{.}{%
}{.}\hspace{.4pt}1145\discretionary{/}{%
}{/}3563657\hspace{.1pt}\discretionary{.}{%
}{.}\hspace{.4pt}3595989}}}


\bibitem{Reinders2020}
S.~Reinders, M.~Butler, and K.~Marriott.
\newblock "hey model!" – natural user interactions and agency in accessible interactive 3d models.
\newblock In {\em Proc. CHI Conference on Human Factors in Computing Systems}, CHI '20,  13 pages, p. 1–13. ACM, New York, 2020. \href{https://doi.org/10.1145/3313831.3376145}
{doi: {{%
10\hspace{.1pt}\discretionary{.}{%
}{.}\hspace{.4pt}1145\discretionary{/}{%
}{/}3313831\hspace{.1pt}\discretionary{.}{%
}{.}\hspace{.4pt}3376145}}}


\bibitem{Reinders2024}
S.~Reinders, M.~Butler, I.~Zukerman, B.~Lee, L.~Qu, and K.~Marriott.
\newblock When refreshable tactile displays meet conversational agents: Investigating accessible data presentation and analysis with touch and speech.
\newblock {\em IEEE Visualization (VIS)}, 2024. \href{https://doi.org/10.48550/arXiv.2408.04806}
{doi: {{%
10\hspace{.1pt}\discretionary{.}{%
}{.}\hspace{.4pt}48550\discretionary{/}{%
}{/}arXiv\hspace{.1pt}\discretionary{.}{%
}{.}\hspace{.4pt}2408\hspace{.1pt}\discretionary{.}{%
}{.}\hspace{.4pt}04806}}}


\bibitem{rowell2003world}
J.~Rowell and S.~Ungar.
\newblock The world of touch: an international survey of tactile maps. part 1: production.
\newblock {\em British Journal of Visual Impairment}, 21(3):98--104, 2003. \href{https://doi.org/10.1177/02646196030210030}
{doi: {{%
10\hspace{.1pt}\discretionary{.}{%
}{.}\hspace{.4pt}1177\discretionary{/}{%
}{/}02646196030210030}}}


\bibitem{Schmitz2012}
B.~Schmitz and T.~Ertl.
\newblock Interactively displaying maps on a tactile graphics display.
\newblock In {\em SKALID 2012 Spatial Knowledge Acquisition with Limited Information Displays}, pp. 13--18, 2012.

\bibitem{SharifEtAlCHI2022}
A.~Sharif, O.~H. Wang, A.~T. Muongchan, K.~Reinecke, and J.~O. Wobbrock.
\newblock Voxlens: Making online data visualizations accessible with an interactive javascript plug-in.
\newblock In {\em Proc. CHI Conference on Human Factors in Computing Systems}, CHI '22,  article no. 478,  19 pages. ACM, New York, 2022. \href{https://doi.org/10.1145/3491102.3517431}
{doi: {{%
10\hspace{.1pt}\discretionary{.}{%
}{.}\hspace{.4pt}1145\discretionary{/}{%
}{/}3491102\hspace{.1pt}\discretionary{.}{%
}{.}\hspace{.4pt}3517431}}}


\bibitem{thompson2023chart}
J.~R. Thompson, J.~J. Martinez, A.~Sarikaya, E.~Cutrell, and B.~Lee.
\newblock Chart reader: Accessible visualization experiences designed with screen reader users.
\newblock In {\em Proc. CHI Conference on Human Factors in Computing Systems}, CHI '23,  article no. 802,  18 pages. ACM, New York, 2023. \href{https://doi.org/10.1145/3544548.3581186}
{doi: {{%
10\hspace{.1pt}\discretionary{.}{%
}{.}\hspace{.4pt}1145\discretionary{/}{%
}{/}3544548\hspace{.1pt}\discretionary{.}{%
}{.}\hspace{.4pt}3581186}}}


\bibitem{Yang2021}
W.~Yang, J.~Huang, R.~Wang, W.~Zhang, H.~Liu, and J.~Xiao.
\newblock A survey on tactile displays for visually impaired people.
\newblock {\em IEEE Transactions on Haptics}, 14(4):712--721, 2021. \href{https://doi.org/10.1109/TOH.2021.3085915}
{doi: {{%
10\hspace{.1pt}\discretionary{.}{%
}{.}\hspace{.4pt}1109\discretionary{/}{%
}{/}TOH\hspace{.1pt}\discretionary{.}{%
}{.}\hspace{.4pt}2021\hspace{.1pt}\discretionary{.}{%
}{.}\hspace{.4pt}3085915}}}


\bibitem{Zeng2015}
L.~Zeng, M.~Miao, and G.~Weber.
\newblock Interactive audio-haptic map explorer on a tactile display.
\newblock {\em Interacting with Computers}, 27(4):413--429, 2015. \href{https://doi.org/10.1093/iwc/iwu006}
{doi: {{%
10\hspace{.1pt}\discretionary{.}{%
}{.}\hspace{.4pt}1093\discretionary{/}{%
}{/}iwc\discretionary{/}{%
}{/}iwu006}}}


\bibitem{zeng2014examples}
L.~Zeng, G.~Weber, I.~Zoller, P.~Lotz, T.~A. Kern, J.~Reisinger, T.~Meiss, T.~Opitz, T.~Rossner, and N.~Stefanova.
\newblock {\em Examples of haptic system development}, pp. 525--554.
\newblock Springer, 2014. \href{https://doi.org/10.1007/978-3-031-04536-3_14}
{doi: {{%
10\hspace{.1pt}\discretionary{.}{%
}{.}\hspace{.4pt}1007\discretionary{/}{%
}{/}978\discretionary{%
}{-}{-}3\discretionary{%
}{-}{-}031\discretionary{%
}{-}{-}04536\discretionary{%
}{-}{-}3\_14}}}


\end{thebibliography}
\end{document}